\documentclass[a4paper,11pt]{article}
\usepackage{pos, xcolor, graphicx, amsmath, xfrac, setspace, physics, slashed, MnSymbol, wasysym}
\usepackage{multirow}
\usepackage{longtable}

\title{Status of the lifetimes of heavy hadrons within the HQE}

\author*[a]{Maria Laura Piscopo}

\affiliation[a]{Physik Department, Universit\"{a}t Siegen, 
Walter-Flex-Str. 3, 57068 Siegen, Germany}

\emailAdd{maria.piscopo@uni-siegen.de}

\abstract{
We present the theoretical status of the lifetimes of weakly decaying heavy hadrons containing a bottom or a charm quark, and discuss the current predictions, based on the framework of the Heavy Quark Expansion (HQE), for both mesons and baryons.\ Potential improvements to reduce the theoretical uncertainties are also highlighted. 

}

\FullConference{%
  8th Symposium on Prospects in the Physics of Discrete Symmetries (DISCRETE 2022)\\
  7-11 November, 2022\\
  Baden-Baden, Germany
}

\begin{document}
\maketitle

\section{Introduction}
The total lifetime $\tau$, or the total decay width $\Gamma = \tau^{-1}$, is one of the fundamental properties of particles.\
In the study of lifetimes, a special role is occupied by heavy hadrons: QCD bound states containing a heavy quark $Q$ with $m_Q \gg \Lambda_{\rm QCD}$, where $\Lambda_{\rm QCD}$ denotes a typical hadronic non-perturbative scale of the order of few hundreds MeV. As their weak decays involve the interplay between the weak and the strong interactions over the wide range of scales $m_W \gg m_Q \gg \Lambda_{\rm QCD}$, heavy hadrons are interesting systems to test the Standard Model (SM) and also, given enough precision in both theory and experiments, perform indirect searches of new physics~(NP).
 
Lifetimes measurements are by now very precise \cite{Workman:2022ynf, HFLAV:2022pwe}.\ On the theoretical side, the Heavy Quark Expansion~(HQE) \cite{Shifman:1984wx} \footnote{See e.g.\ the review \cite{Lenz:2014jha} for a more comprehensive list of references.} provides a well established framework to compute inclusive decay widths of heavy hadrons in terms of a systematic expansion in inverse powers of $m_Q$.\ As the reliability of the HQE strongly depends on the assumption that $Q$ is heavy, bottom hadrons are clearly the most suited systems to be described within this framework.\ The charm quark mass, on the other hand, lies at the boundary between the heavy and the light quark regime, and if charmed systems pose notoriously more challenges to precise theoretical studies, they also give an opportunity to test even further the applicability of the HQE.
\section{The HQE in the bottom sector}
The total decay width of a bottom hadron $H_b$ is given by
\begin{equation} 
\Gamma(H_b) = \frac{1}{2 m_{H_b}} \sum_n \int_{\rm PS} (2 \pi)^4 \delta^{(4)} (p_n- p_{H_b}) |\langle n| {\cal H}_{\rm eff}| H_b \rangle |^2\,,
\label{eq:Gamma-def}
\end{equation}
where the sum runs over all possible final states into which $H_b$ can decay, ${\rm PS}$ is the corresponding phase space, and ${\cal H}_{\rm eff}$ the effective Hamiltonian \cite{Buchalla:1995vs}. Using the optical theorem, eq.~\eqref{eq:Gamma-def} becomes
\begin{equation}
\Gamma(H_b) = \frac{1}{2 m_{H_b}} {\rm Im} \langle H_b| {\cal T} |  H_b \rangle \,, 
\quad 
{\rm with}
\quad 
{\cal T} = i  \!\! \int \!\! d^4x  \, \, {\rm T} \{ {\cal H}_{\rm eff}(x), {\cal H}_{\rm eff}(0)  \} \,.
\label{eq:calT}
\end{equation}
The transition operator ${\cal T}$ contains the time-ordered product of ${\cal H}_{\rm eff}$ and is thus a non-local operator; its expression however, simplifies in the limit of a heavy decaying quark.\ In fact, by exploiting the hierarchy $m_b \gg \Lambda_{\rm QCD}$, the $b$-quark momentum and field can be respectively parametrised as 
\begin{equation}
p_b^\mu = m_b v^\mu + k^\mu\,,   
\qquad 
b(x) = e^{- i m_b v \cdot x} b_v (x)\,, 
\label{eq:HQE-p}
\end{equation}
where $v = p_{H_b}/m_{H_b}$ is the hadron velocity, $k \ll m_b$ a residual momentum of the order of $\Lambda_{\rm QCD}$, and we have introduced the rescaled field $b_v(x)$ containing only low oscillation frequencies of the order of $k$.\
\begin{table}[th]
\centering
\renewcommand{\arraystretch}{1.3}
\begin{tabular}{|c|c|}
\hline
\multicolumn{2}{|c|}{Semileptonic modes (SL)} \\
\hline 
\multirow{2}{*}{$\Gamma_3^{(3)}$} & 
{\scriptsize \it Fael, Sch\"onwald, Steinhauser \cite{Fael:2020tow}} \\[-2mm]
& 
{\scriptsize \it Czakon, Czarnecki, Dowling \cite{Czakon:2021ybq}} \\
\hline
\multirow{2}{*}{$\Gamma_5^{(1)}$} & 
 {\scriptsize \it  Alberti, Gambino, Nandi \cite{Alberti:2013kxa}} \\[-2mm]
& {\scriptsize \it  Mannel, Pivovarov, Rosenthal \cite{Mannel:2015jka}}\\
\hline
\multirow{1}{*}{$\Gamma_6^{(1)}$} &
 {\scriptsize \it  Mannel, Moreno, Pivovarov \cite{Mannel:2019qel}}
 \\
\hline
\multirow{1}{*}{$\Gamma_7^{(0)}$} & 
{\scriptsize \it  Dassinger, Mannel, Turczyk \cite{Dassinger:2006md}} \\
\hline
\multirow{1}{*}{$\Gamma_8^{(0)}$} & 
 {\scriptsize \it  Mannel, Turczyk, Uraltsev \cite{Mannel:2010wj}} \\
\hline
\multirow{1}{*}{$\tilde \Gamma_6^{(1)}$} & 
{\scriptsize \it  Lenz, Rauh \cite{Lenz:2013aua}} \\
\hline
\end{tabular}
\begin{tabular}{|c|c|}
\hline
\multicolumn{2}{|c|}{Non-leptonic modes (NL)} \\
\hline 
\multirow{1}{*}{$\Gamma_3^{(2)}$} &  
{\scriptsize \it  Czarnecki, Slusarczyk, Tkachov \cite{Czarnecki:2005vr}$^*$} \\
\hline 
\multirow{2}{*}{$\Gamma_3^{(1)}$} &  
{\scriptsize \it  Ho-Kim, Pham \cite{Ho-kim:1983klw};  Altarelli, Petrarca \cite{Altarelli:1991dx}; Bagan et al.\ \cite{Bagan:1994zd}} \\[-2mm]
& {\scriptsize \it   
Lenz, Nierste, Ostermaier \cite{Lenz:1997aa}; Krinner,  Lenz, Rauh \cite{Krinner:2013cja}}\\
\hline
\multirow{1}{*}{$\Gamma_5^{(0)}$} &  
{\scriptsize \it  Bigi, Uraltsev, Vainsthtein \cite{Bigi:1992su}; Blok, Shifman \cite{Blok:1992he}}
\\
\hline
\multirow{1}{*}{$\Gamma_6^{(0)}$} &  
{\scriptsize \it  Lenz, Piscopo, Rusov \cite{Lenz:2020oce}; Mannel, Moreno, Pivovarov \cite{Mannel:2020fts}}
 \\
\hline
\multirow{2}{*}{$\tilde \Gamma_6^{(1)}$} &  
{\scriptsize \it Beneke, Buchalla, Greub, Lenz, Nierste \cite{Beneke:2002rj}} \\[-2mm]
& {\scriptsize \it  Franco, Lubicz, Mescia, Tarantino  \cite{Franco:2002fc}} \\
\hline
\multirow{1}{*}{$\tilde \Gamma_7^{(0)}$} & 
 {\scriptsize \it Gabbiani, Onishchenko, Petrov \cite{Gabbiani:2003pq}} \\
\hline
\end{tabular}
\caption{Status of the perturbative corrections in the HQE. The work with $^*$ contains only partial results.}
\label{tab:Status1}
\end{table} 
Taking into account eq.~\eqref{eq:HQE-p}, the time ordered product in eq.~\eqref{eq:calT} can be systematically expanded into a series of local operators ${\cal O}_d$ of increasing dimension $d$ and suppressed by $d-3$ powers of the heavy quark mass. In this way, the total decay width takes the following form
\begin{equation}
\Gamma(H_b) = 
 \Gamma_3  +
 \Gamma_5  \frac{\langle  {\cal O}_5  \rangle}{m_b^2} + 
 \Gamma_6 \frac{\langle  {\cal O}_6  \rangle}{m_b^3} + \ldots
 + 16 \pi^2 
\left[ 
 \tilde{\Gamma}_6 \frac{\langle \tilde{\mathcal{O}}_6 \rangle}{m_b^3} +
 \tilde{\Gamma}_7 \frac{\langle \tilde{\mathcal{O}}_7 \rangle}{m_b^4} + \dots 
\right]\,.
\label{eq:HQE}
\end{equation}
The short-distance functions $\Gamma_d$ can be computed in terms of a perturbative expansion in $\alpha_s(m_b)$
\begin{equation}
\Gamma_d = \Gamma_d^{(0)} + \left( \frac{\alpha_s}{4 \pi}\right) \Gamma_d^{(1)} +  \left( \frac{\alpha_s}{4 \pi}\right)^2 \Gamma_d^{(2)} + \ldots \,,
\label{eq:HQE-per}
\end{equation}
whereas $\langle {\cal O}_d \rangle \equiv \langle H_b| {\cal O}_d |H_b \rangle /(2 m_{H_b})$ are matrix elements parametrising the non-perturbative effects. Note that in eq.~\eqref{eq:HQE} both two- and four-quark operators contribute.\ While at LO-QCD the former are obtained from the discontinuity of two-loop diagrams, the latter already arise at one-loop at the same order in $\alpha_s$,
hence the explicit phase space factor $16 \pi^2$ in front of the four-quark operators contributions, here labeled by a tilde.
Lifetime ratios of bottom hadrons can then be computed as
\begin{equation}
  \frac{\tau(H_b)}{\tau(H_{b}^\prime)}  =  1 + \left[ \Gamma (H_{b}^\prime)^{\, \rm HQE} - \Gamma (H_b)^{\, \rm HQE}\right]  \tau(H_b)^{\rm exp.} \, ,
  \label{eq:ratio1}
\end{equation}
where the experimental value of $\tau(H_b)$ is used to cancel out the dependence on $\Gamma_3$, albeit this is
not necessary and eq.~\eqref{eq:ratio1} can be also computed entirely within the HQE with only slightly larger uncertainties, see e.g.\ the discussion in \cite{Lenz:2022rbq}. It is worth emphasising that while two-quark operators present short-distance coefficients which are universal for a given heavy quark, so that any differences in lifetimes of bottom hadrons would only be induced by the value of their matrix elements, the Wilson coefficients of the four-quark operators also depend on the spectator quark in the specific hadron considered and this can in general lead to larger effects in lifetime ratios.

The HQE is by now a very advanced framework. The theoretical status for both the perturbative and non-perturbative side is summarised respectively in tables~\ref{tab:Status1} and \ref{tab:Status2}.

\begin{table}
\centering
\renewcommand{\arraystretch}{1.7}
\begin{tabular}{|c||c|c||c|c|}
\hline
& $B_d, B^+$ & $B_s$  & $\Lambda^0_b$  & $\Xi_b^-$, $\Xi_b^0$, $\Omega_b^- $ \\
\hline 
\multirow{3}{*}{ $\langle {\cal O}_5 \rangle $} &  
{\scriptsize \it Fits to SL data \cite{Alberti:2014yda, Gambino:2016jkc, Bordone:2021oof, Bernlochner:2022ucr} }   & 
& 
&
\\[-3mm]
& {\scriptsize \it  HQET sum rules \cite{Ball:1993xv, Neubert:1996wm} }  &  
{\scriptsize \it Spectroscopy \cite{Bigi:2011gf} } &  
{\scriptsize \it Spectroscopy \cite{Bigi:1992su}}
&
{\scriptsize \it Spectroscopy  \cite{Gratrex:2022xpm}}
 \\[-3mm]
& {\scriptsize \it  Lattice QCD \cite{Gambino:2017vkx, FermilabLattice:2018est} } &
{\scriptsize \it  } 
&
&  
\\
\hline
\multirow{2}{*}{$\langle {\cal O}_6 \rangle$}  & 
{\scriptsize \it Fits to SL data \cite{Alberti:2014yda, Gambino:2016jkc, Bordone:2021oof, Bernlochner:2022ucr} } & 
{\scriptsize \it Sum rules estimates \cite{Bigi:2011gf}} &
{\scriptsize \it EOM relation to $\langle \tilde {\cal O}_6 \rangle$} 
&
 {\scriptsize \it EOM relation to $\langle \tilde {\cal O}_6 \rangle$}
 \\[-3mm]
 & 
 {\scriptsize \it EOM relation to $\langle \tilde {\cal O}_6 \rangle$} &
 {\scriptsize \it EOM relation to $\langle \tilde {\cal O}_6 \rangle$}
 & 
 &
 \\ 
 \hline
\multirow{2}{*}{$\langle \tilde {\cal O}_6 \rangle$} &  
{\scriptsize \it HQET sum rules \cite{Kirk:2017juj}} &
{\scriptsize\it HQET sum rules \cite{King:2021jsq}} &
{\scriptsize \it HQET SR \cite{Colangelo:1996ta}; NRCQM + }
&
{\scriptsize \it NRCQM + }
\\[-3mm]
& & & 
{\scriptsize \it spectroscopy \cite{DeRujula:1975qlm, Rosner:1996fy} }
&
{\scriptsize \it spectroscopy \cite{DeRujula:1975qlm, Rosner:1996fy} }
\\
\hline
$\langle \tilde {\cal O}_7 \rangle$ &
\multicolumn{2}{|c||}{\scriptsize \it Vacuum insertion approximation} 
&
-----
&
----
\\ 
\hline
\end{tabular} 
\caption{Status of the available determinations of the non-perturbative matrix elements in the HQE for the bottom sector. The following abbreviations are used: heavy quark effective theory (HQET), sum rules (SR), equation of motion (EOM), non-relativistic constituent quark model (NRCQM).}
\label{tab:Status2}
\end{table}
\subsection{Lifetimes of bottom mesons}
\begin{figure}[h]
    \centering
    \includegraphics[scale=0.8]{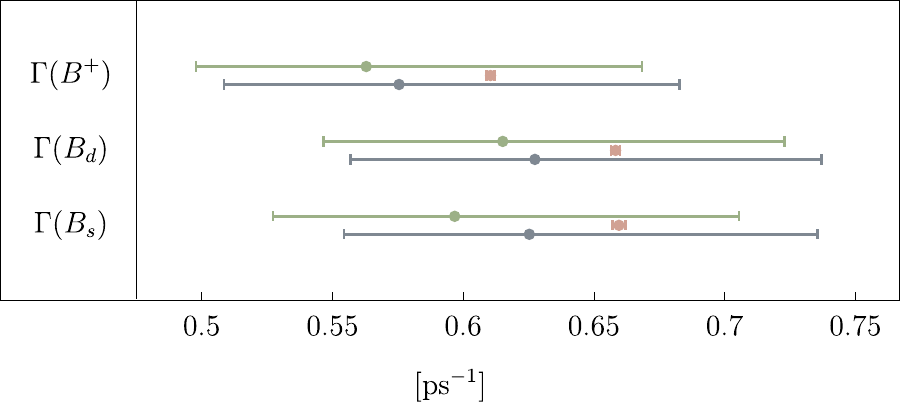}
    \includegraphics[scale=0.8]{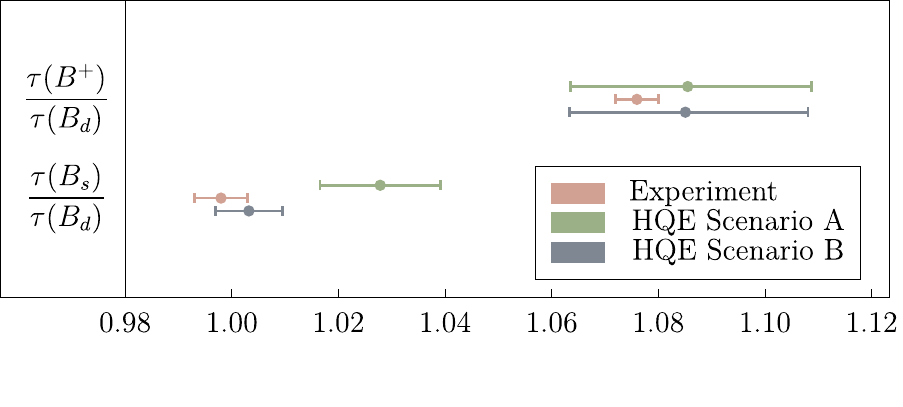}
    \caption{Comparison of data and HQE predictions for two different choices of theory inputs, based on \cite{Lenz:2022rbq}.}
    \label{fig:B-mesons-results}
\end{figure}

The current HQE predictions for the total widths of the bottom mesons $B_d, B^+, B_s,$ and their lifetime ratios, are shown in fig.~\ref{fig:B-mesons-results} and table~\ref{tab:summary-with-uncertainties} \cite{Lenz:2022rbq}.\ The two scenarios correspond to different choices of inputs, including the value of the $b$-quark mass and of the non-perturbative parameters associated to the matrix elements of two-quark operators, see for details \cite{Lenz:2022rbq}.\ While the total decay widths and the ratio $\tau(B^+)/\tau(B_d)$
are only mildly sensitive to this choice, the results for $\tau(B_s)/\tau(B_d)$ are strongly affected, and in one case, a small tension with the experimental value emerges.\ In this respect, we point out that the two fits to experimental data on inclusive semileptonic $B$-meson decays~\cite{Bordone:2021oof, Bernlochner:2022ucr} \footnote{We stress that corresponding measurements for inclusive semileptonic $B_s$-decays are still missing.} find different values of the Darwin parameter $\rho_D^3$, which plays a crucial role in the prediction of $\tau(B_s)/\tau(B_d)$.\ Further insights on the origin of this discrepancy are clearly of utmost importance given the potential that this observable could have as an indirect probe of NP. Moreover, an investigation by lattice QCD of the small size of  the bag-parameters of the octet-operators and of the so-called `eye-contractions' as found in \cite{Kirk:2017juj, King:2021jsq}, would also be very desirable.

Scale variation in the leading term $\Gamma_3$, and the values of the non-perturbative parameters, including SU(3)$_F$ breaking effects, are the main sources of uncertainties for the total widths and the lifetime ratios, respectively. Overall, there is very good agreement between HQE and data,
and the clean observable $\tau(B^+)/\tau(B_d)$ could already be used to constrain certain NP operators~\cite{Lenz:2022pgw}.
\begin{table}[h]\centering
\renewcommand{\arraystretch}{1.7}
\begin{tabular}{|c||c|c||c|}
\hline
& HQE Scenario A  &  HQE Scenario B & Exp. value \\
\hline
\hline
$\tau(B^+)/\tau(B_d) $ 
& $1.0855^{+0.0232}_{-0.0219} $
& $1.0851^{+0.0230}_{-0.0217} $
& $1.076 \pm 0.004 $
\\
\hline
$\tau(B_s)/\tau(B_d)$ 
& $1.0279^{+0.0113}_{-0.0113} $
& $1.0032^{+0.0063}_{-0.0063} $
& $0.998 \pm 0.005 $
\\
\hline
\end{tabular}
\caption{HQE predictions of the lifetime ratios for two different choices of theory inputs, see \cite{Lenz:2022rbq}.}
\label{tab:summary-with-uncertainties}
\end{table}
\subsection{Lifetimes of bottom baryons}
\begin{figure}[ht]
    \centering
    \includegraphics[scale=0.8]{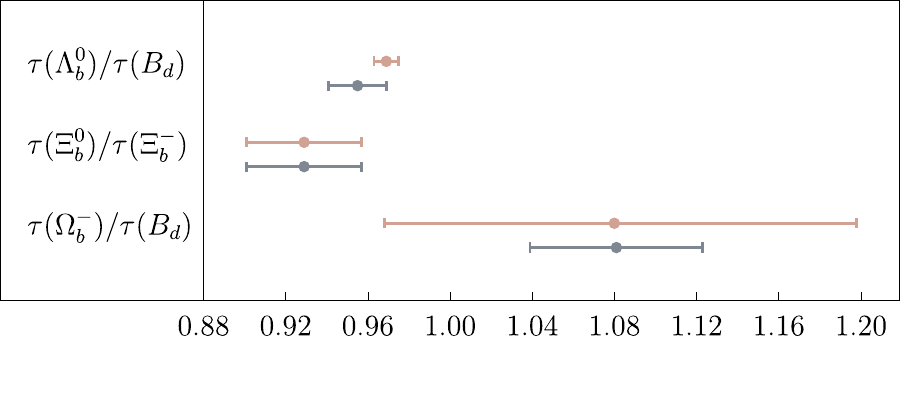} 
    \includegraphics[scale=0.8]{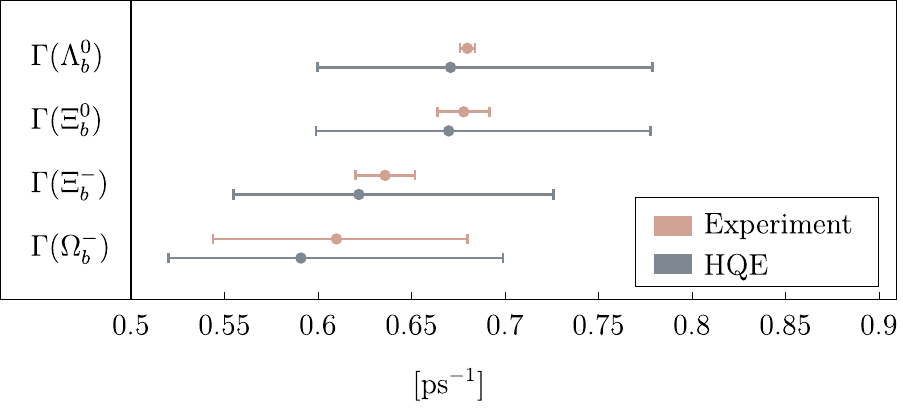}
    \caption{Comparison of HQE predictions and experimental data for $b$-baryons \cite{Gratrex:2023pfn}.}
    \label{fig:HQE-vs-Data}
\end{figure}

The most recent theoretical predictions for the total widths of the $b$-baryons $\Lambda^0_b, \Xi_b^0, \Xi_b^-, \Omega_b^-,$ as well as their lifetime ratios, are displayed in fig.~\ref{fig:HQE-vs-Data} and in table~\ref{tab:exp-data}, see \cite{Gratrex:2023pfn}.\ Along all the observables considered there is very good agreement between theory and experiments, while the dominant sources of uncertainties are again scale variation and the values of the non-perturbative parameters.\ In fact, despite the efforts in e.g.~\cite{Colangelo:1996ta, DiPierro:1999tb}, the four-quark dimension-six matrix-elements have not yet been computed for all baryons either within lattice QCD or QCD sum rules, and currently only spectroscopy relations based on simplified models of QCD, like the NRCQM, can be consistently used.\ First principle calculations would be of great importance in order to reduce the overall uncertainties and obtain an independent determination of these non-perturbative parameters.
\begin{table}[h]
\centering
\renewcommand{\arraystretch}{1.7}
    \begin{tabular}{|c||c|c|c|}
    \hline
         & $\tau(\Lambda^0_b)/\tau(B_d) $
         & $\tau(\Xi_b^0)/\tau(\Xi_b^-)$ 
         & $\tau(\Omega^-_b)/\tau(B_d)$
         \\
\hline
\hline
    HQE
    & $0.955 \pm 0.014 $
    & $0.929 \pm 0.028 $
    & $1.081 \pm 0.042 $
    \\
    \hline
    \hline
    Exp. 
     & $0.969 \pm 0.006 $
     & $0.929 \pm 0.028 $
     & $1.080^{+0.118}_{-0.112}$
    \\
    \hline
    \end{tabular}
    \caption{
    Comparison of data and HQE predictions for selected lifetime ratios, based on \cite{Gratrex:2023pfn}.
   }
    \label{tab:exp-data}
\end{table}
\section{The HQE at its limit: lifetimes of charmed hadrons}
While the applicability of the HQE appears to be well established in the bottom sector, it is questionable whether it also extends to the charm system, given that $m_c \sim 1$ GeV. In this case the two expansion parameters $\alpha_s(m_c)$ and $\Lambda_{\rm QCD}/m_c$ become larger, making both the perturbative and the power-correction series a priori less reliable.\ Furthermore, many of the non-perturbative parameters of the HQE for charmed hadrons are still poorly known and in most of the cases only rough estimates based on symmetry arguments with respect to the bottom sector are available.  

\begin{figure}[h]
    \centering
    \includegraphics[scale=0.8]{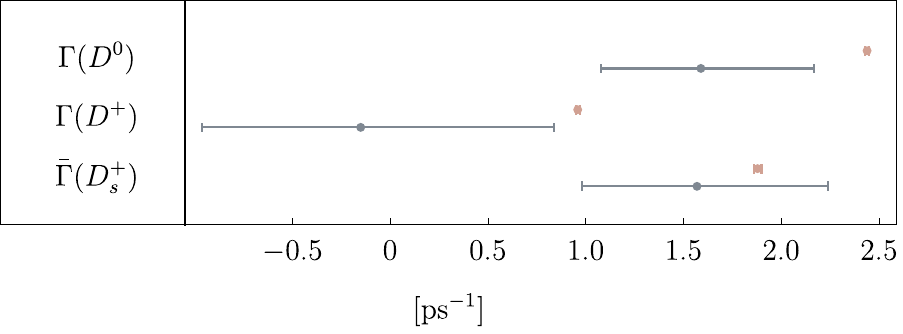} \,
    \includegraphics[scale=0.8]{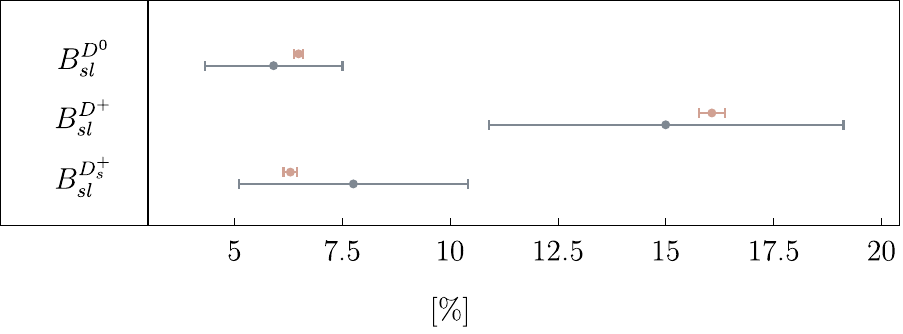} \\
    \includegraphics[scale=0.8]{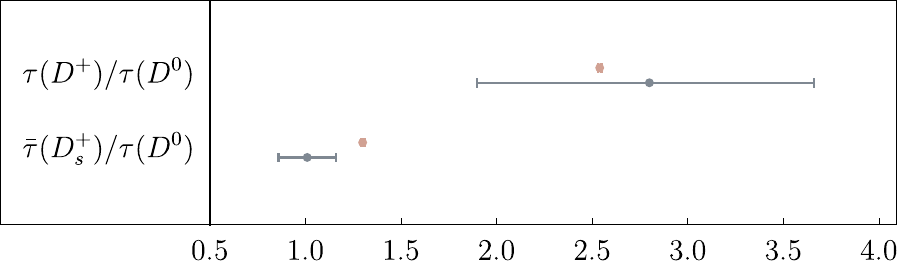} \,
    \includegraphics[scale=0.8]{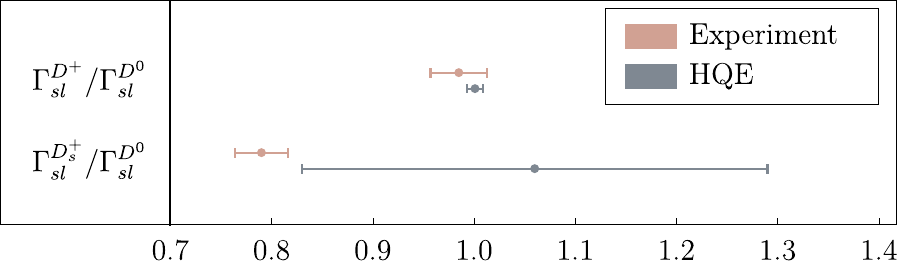}
    \caption{Comparison of HQE predictions and experimental data for charmed mesons, based on \cite{King:2021xqp}.}
    \label{fig:summary-comparison-charm}
\end{figure}

It is then interesting to confront the HQE framework with the experimental data for the charm system.\ The current theoretical predictions for the total widths of the charmed mesons $D^0, D^+, D_s^+$, their lifetime ratios, as well as the semileptonic branching fractions, are shown in fig.~\ref{fig:summary-comparison-charm} and table~\ref{tab:exp-data}~\cite{King:2021xqp}.\ Although with very large uncertainties, the HQE succeeds in correctly describing the observed patterns, and no indication of a possible breakdown of the framework seems to emerge.\ These results have been confirmed in the recent study \cite{Gratrex:2022xpm}, which included as well predictions for the lifetimes of singly charmed baryons.\ Also in this case, it is found that the HQE can consistently accommodate the experimental data, albeit again within large uncertainties.
\begin{table}[h]
\centering
\renewcommand{\arraystretch}{1.7}
    \begin{tabular}{|c||c|c|}
    \hline
         & $\tau(D^+)/\tau(D^0)$ 
         & $\bar \tau(D_s^+)/\tau(D^0)$
         \\
\hline
\hline
    HQE
    & $2.80^{+0.86}_{-0.90} $
    & $ 1.00 \pm 0.16$
    \\
    \hline
    \hline
    Exp. 
     & $2.54 \pm 0.02 $
     & $1.30 \pm 0.01 $
    \\
    \hline
    \end{tabular}
    \caption{Comparison of data and HQE predictions of charmed mesons lifetime ratios, based on \cite{King:2021xqp}. Note that $\bar \tau(D_s^+)$ does not include the semileptonic mode $D_s^+ \to \tau^+ \nu_\tau$, see for details \cite{King:2021xqp}. }
    \label{tab:exp-data}
\end{table}
\section{Conclusions}
We have presented the status of the lifetimes of heavy hadrons containing a bottom or charm quark within the HQE \footnote{We have not discussed hadrons containing two heavy quarks. For studies of the $B_c$ lifetime see \cite{Beneke:1996xe, Aebischer:2021ilm}.}. Along all the observables considered the agreement with the experimental data is good, confirming the HQE as a powerful framework for inclusive heavy hadron decays, contrary to some conclusions in \cite{Cheng:2018rkz}.\ Moreover, the precision already achieved for several clean quantities in the bottom system is high and comparable with the experimental one.\ Further improvements in the computation of higher order corrections, namely $\Gamma_{3,{\rm NL}}^{(2)}$, $\Gamma_7^{(0)}$, $\tilde \Gamma_6^{(2)}$, as well as in the determination of the non-perturbative parameters,
will be crucial in order to reduce the theoretical uncertainties, so that in future these observables might also serve as interesting probes of~NP.
\section{Acknowledgments}
I would like to thank the organisers of DISCRETE 2022 for the invitation, and Alexander Lenz and Aleksey Rusov for helpful comments on the manuscript. This work is funded by the Deutsche Forschungsgemeinschaft (DFG, German Research Foundation) - project number 500314741.


\begin{thebibliography}{99}

\bibitem{Workman:2022ynf}
R.~L.~Workman \textit{et al.} [Particle Data Group],
PTEP \textbf{2022} (2022), 083C01.

\bibitem{HFLAV:2022pwe}
Y.~S.~Amhis \textit{et al.} [HFLAV],
[arXiv:2206.07501 [hep-ex]].

\bibitem{Shifman:1984wx}
M.~A.~Shifman and M.~B.~Voloshin,
Sov. J. Nucl. Phys. \textbf{41} (1985), 120.

\bibitem{Lenz:2014jha}
A.~Lenz,
Int. J. Mod. Phys. A \textbf{30} (2015) no.10, 1543005.

\bibitem{Buchalla:1995vs}
G.~Buchalla, A.~J.~Buras and M.~E.~Lautenbacher,
Rev. Mod. Phys. \textbf{68} (1996), 1125-1144.

\bibitem{Fael:2020tow}
M.~Fael, K.~Sch\"onwald and M.~Steinhauser,
Phys. Rev. D \textbf{104} (2021) no.1, 016003.

\bibitem{Czakon:2021ybq}
M.~Czakon, A.~Czarnecki and M.~Dowling,
Phys. Rev. D \textbf{103} (2021), L111301.

\bibitem{Alberti:2013kxa}
A.~Alberti, P.~Gambino and S.~Nandi,
JHEP \textbf{01} (2014), 147.

\bibitem{Mannel:2015jka}
T.~Mannel, A.~A.~Pivovarov and D.~Rosenthal,
Phys. Rev. D \textbf{92} (2015) no.5, 054025.

\bibitem{Mannel:2019qel}
T.~Mannel and A.~A.~Pivovarov,
Phys. Rev. D \textbf{100} (2019) no.9, 093001. 
T.~Mannel, D.~Moreno and A.~A.~Pivovarov,
Phys. Rev. D \textbf{105} (2022) no.5, 054033.
D.~Moreno,
Phys. Rev. D \textbf{106} (2022) no.11, 114008.

\bibitem{Dassinger:2006md}
B.~M.~Dassinger, T.~Mannel and S.~Turczyk,
JHEP \textbf{03} (2007), 087.

\bibitem{Mannel:2010wj}
T.~Mannel, S.~Turczyk and N.~Uraltsev,
JHEP \textbf{11} (2010), 109.

\bibitem{Lenz:2013aua}
A.~Lenz and T.~Rauh,
Phys. Rev. D \textbf{88} (2013), 034004.

\bibitem{Czarnecki:2005vr}
A.~Czarnecki, M.~Slusarczyk and F.~V.~Tkachov,
Phys. Rev. Lett. \textbf{96} (2006), 171803.

\bibitem{Ho-kim:1983klw}
Q.~Ho-kim and X.~Y.~Pham,
Annals Phys. \textbf{155} (1984), 202.

\bibitem{Altarelli:1991dx}
G.~Altarelli and S.~Petrarca,
Phys. Lett. B \textbf{261} (1991), 303-310.

\bibitem{Bagan:1994zd}
E.~Bagan, P.~Ball, V.~M.~Braun and P.~Gosdzinsky,
Nucl. Phys. B \textbf{432} (1994), 3-38.

\bibitem{Lenz:1997aa}
A.~Lenz, U.~Nierste and G.~Ostermaier,
Phys. Rev. D \textbf{56} (1997), 7228-7239; Phys. Rev. D \textbf{59} (1999), 034008.

\bibitem{Krinner:2013cja}
F.~Krinner, A.~Lenz and T.~Rauh,
Nucl. Phys. B \textbf{876} (2013), 31-54.

\bibitem{Bigi:1992su}
I.~I.~Y.~Bigi, N.~G.~Uraltsev and A.~I.~Vainshtein,
Phys. Lett. B \textbf{293} (1992), 430-436.

\bibitem{Blok:1992he}
B.~Blok and M.~A.~Shifman,
Nucl. Phys. B \textbf{399} (1993), 441-458;
Nucl. Phys. B \textbf{399} (1993), 459-476.

\bibitem{Lenz:2020oce}
A.~Lenz, M.~L.~Piscopo and A.~V.~Rusov,
JHEP \textbf{12} (2020), 199.

\bibitem{Mannel:2020fts}
T.~Mannel, D.~Moreno and A.~Pivovarov,
JHEP \textbf{08} (2020), 089.  D.~Moreno,
JHEP \textbf{01} (2021), 051.

\bibitem{Beneke:2002rj}
M.~Beneke, G.~Buchalla, C.~Greub, A.~Lenz and U.~Nierste,
Nucl. Phys. B \textbf{639} (2002), 389-407.

\bibitem{Franco:2002fc}
E.~Franco, V.~Lubicz, F.~Mescia and C.~Tarantino,
Nucl. Phys. B \textbf{633} (2002), 212-236.

\bibitem{Gabbiani:2003pq}
F.~Gabbiani, A.~I.~Onishchenko and A.~A.~Petrov,
Phys. Rev. D \textbf{68} (2003), 114006; Phys. Rev. D \textbf{70} (2004), 094031.

\bibitem{Lenz:2022rbq}
A.~Lenz, M.~L.~Piscopo and A.~V.~Rusov,
JHEP \textbf{01} (2023), 004.

\bibitem{Alberti:2014yda}
A.~Alberti, P.~Gambino, K.~J.~Healey and S.~Nandi,
Phys. Rev. Lett. \textbf{114} (2015) no.6, 061802.

\bibitem{Gambino:2016jkc}
P.~Gambino, K.~J.~Healey and S.~Turczyk,
Phys. Lett. B \textbf{763} (2016), 60-65.

\bibitem{Bordone:2021oof}
M.~Bordone, B.~Capdevila and P.~Gambino,
Phys. Lett. B \textbf{822} (2021), 136679.

\bibitem{Bernlochner:2022ucr}
F.~Bernlochner, M.~Fael, K.~Olschewsky, E.~Persson, R.~van Tonder, K.~K.~Vos and M.~Welsch,
JHEP \textbf{10} (2022), 068.

\bibitem{Ball:1993xv}
P.~Ball and V.~M.~Braun,
Phys. Rev. D \textbf{49} (1994), 2472-2489.

\bibitem{Neubert:1996wm}
M.~Neubert,
Phys. Lett. B \textbf{389} (1996), 727-736.

\bibitem{Gambino:2017vkx}
P.~Gambino, A.~Melis and S.~Simula,
Phys. Rev. D \textbf{96} (2017) no.1, 014511.

\bibitem{FermilabLattice:2018est}
A.~Bazavov \textit{et al.}, 
Phys. Rev. D \textbf{98} (2018) no.5, 054517.

\bibitem{Kirk:2017juj}
M.~Kirk, A.~Lenz and T.~Rauh,
JHEP \textbf{12} (2017), 068.

\bibitem{Bigi:2011gf}
I.~I.~Bigi, T.~Mannel and N.~Uraltsev,
JHEP \textbf{09} (2011), 012.

\bibitem{King:2021jsq}
D.~King, A.~Lenz and T.~Rauh,
JHEP \textbf{06} (2022), 134.

\bibitem{Colangelo:1996ta}
P.~Colangelo and F.~De Fazio,
Phys. Lett. B \textbf{387} (1996), 371-378.

\bibitem{DiPierro:1999tb}
M.~Di Pierro \textit{et al.} [UKQCD],
Phys. Lett. B \textbf{468} (1999), 143.

\bibitem{DeRujula:1975qlm}
A.~De Rujula, H.~Georgi and S.~L.~Glashow,
Phys. Rev. D \textbf{12} (1975), 147-162.

\bibitem{Rosner:1996fy}
J.~L.~Rosner,
Phys. Lett. B \textbf{379} (1996), 267-271.

\bibitem{Gratrex:2022xpm}
J.~Gratrex, B.~Meli\'c and I.~Ni\v{s}and\v{z}i\'c,
JHEP \textbf{07} (2022), 058.

\bibitem{Lenz:2022pgw}
A.~Lenz, J.~M\"uller, M.~L.~Piscopo and A.~V.~Rusov,
[arXiv:2211.02724 [hep-ph]].

\bibitem{Gratrex:2023pfn}
J.~Gratrex, A.~Lenz, B.~Meli\'c, I.~Ni\v{s}and\v{z}i\'c, M.~L.~Piscopo and A.~V.~Rusov,~[arXiv:2301.07698].

\bibitem{King:2021xqp}
D.~King, A.~Lenz, M.~L.~Piscopo, T.~Rauh, A.~V.~Rusov and C.~Vlahos,
JHEP \textbf{08} (2022), 241.

\bibitem{Beneke:1996xe}
M.~Beneke and G.~Buchalla,
Phys. Rev. D \textbf{53} (1996), 4991-5000.

\bibitem{Aebischer:2021ilm}
J.~Aebischer and B.~Grinstein,
JHEP \textbf{07} (2021), 130; Phys. Lett. B \textbf{834} (2022), 137435.

\bibitem{Cheng:2018rkz}
H.~Y.~Cheng,
JHEP \textbf{11} (2018), 014.



\end{thebibliography}
\end{document}